\documentclass[12pt]{article}
\usepackage{latexsym}
\usepackage{epsfig,amssymb,euscript}
\usepackage{amsmath}
\usepackage{array,calc,epsfig}

\def\be{\begin{equation}}
\def\ee{\end{equation}}
\def\bea{\begin{eqnarray}}
\def\eea{\end{eqnarray}}

\begin{document}
\baselineskip=15.5pt
\pagestyle{plain}
\setcounter{page}{1}
\begin{titlepage}

\leftline{\tt hep-th/0611253}

\vskip -.8cm

\rightline{\small{\tt UB-ECM-PF 06/39}}
\rightline{\small{\tt KUL-TF-06/32}}

\begin{center}

\vskip 1.4 cm

{\LARGE{\bf Meson decays from string splitting}}
\vskip .3cm 

\vskip 1.2cm
\vspace{20pt}
{\large 
F. Bigazzi $^{a}$, A. L. Cotrone $^{b}$, L. Martucci$^{c}$, W. Troost$^{d}$}\\
\vskip 1.2cm

\textit{$^a$ LPTHE, Universit\'es Paris VI and VII, 4 place Jussieu; 75005, Paris,
France, and\\
Physique Th\'eorique et Math\'ematique and International Solvay
Institutes, Universit\'e Libre de Bruxelles; CP 231, B-1050 Bruxelles,
Belgium.
}\\
\textit{$^b$ Departament ECM, Facultat de F\'isica, Universitat de Barcelona and \\ Institut
de Fisica d'Altes Energies, Diagonal 647, E-08028 Barcelona, Spain.}\\
\textit{$^c$ Institute for theoretical physics, K.U. Leuven,\\
Celestijnenlaan 200D, B-3001 Leuven, Belgium.}\\
\end{center}

\vspace{12pt}

\begin{center}
\textbf{Abstract}
\end{center}

\vspace{4pt}{\small \noindent We discuss exclusive decays of large spin mesons into
mesons in models of large $N_c$ quenched QCD at strong coupling using string
theory. The rate of the processes are calculated by studying the splitting
of a macroscopic string on the relevant dual gravity backgrounds.
We study analytic formulas for the decay rates of mesons
made up of very heavy or very light quarks.
}

\vfill \vskip 5.mm \hrule width 5.cm \vskip 2.mm {\small \noindent e-mails: bigazzi@lpthe.jussieu.fr, cotrone@ecm.ub.es, luca.martucci@fys.kuleuven.be,\\ walter.troost@fys.kuleuven.be}

\noindent
\end{titlepage}

\newpage


\section{Introduction}

The String/Gauge theory correspondence states the equivalence of a string theory (or M-theory) on a non-trivial background and a gauge theory.
Once a supergravity background dual to some gauge theory with adjoint fields is given, the easiest way to add fundamental flavors is by placing $N_f$ ``flavor branes'' in the background \cite{kk}. 
In the probe approximation $N_f\ll N_c$ ($N_c$ being the number of colors) we can just ignore the backreaction of the flavor branes.
In this way it is quite easy to study meson spectra:
small brane fluctuations correspond to small spin mesons, while
macroscopic spinning strings attached to the branes describe large spin mesons.

Compared to spectral properties, dynamical processes in the string/gauge theory duality are much less studied (relevant exceptions are in \cite{ps}).
Here we present one of the few examples, by studying decays of high spin mesons that have a description as string splitting.
In particular, we study certain exclusive decays of high spin mesons into mesons in models of large $N_c$ quenched QCD at strong coupling \cite{noi1,noi2}.

\section{Two ``models of QCD''}

The setting of our investigation are two string models that are similar to QCD. 
Both are based on the same supergravity background, generated by $N_c$ D4-branes wrapped on a supersymmetry breaking circle \cite{witten}. 
At low energies this gives a four dimensional large $N_c$ Yang-Mills theory, that in the regime where supergravity is reliable is coupled to a tower of adjoint KK fields.

In the first model (that we will call Model I) one adds to this background $N_f\ll N_c$ D6-branes, whose position at minimal radius is connected to the quark mass $m_Q$ \cite{myers}.
In the large spin $J$ regime the string that describes the meson is almost ``straight'' \cite{pando}, see Figure 1 (a).
\begin{figure}
\begin{center}
\scalebox{0.7}{\includegraphics{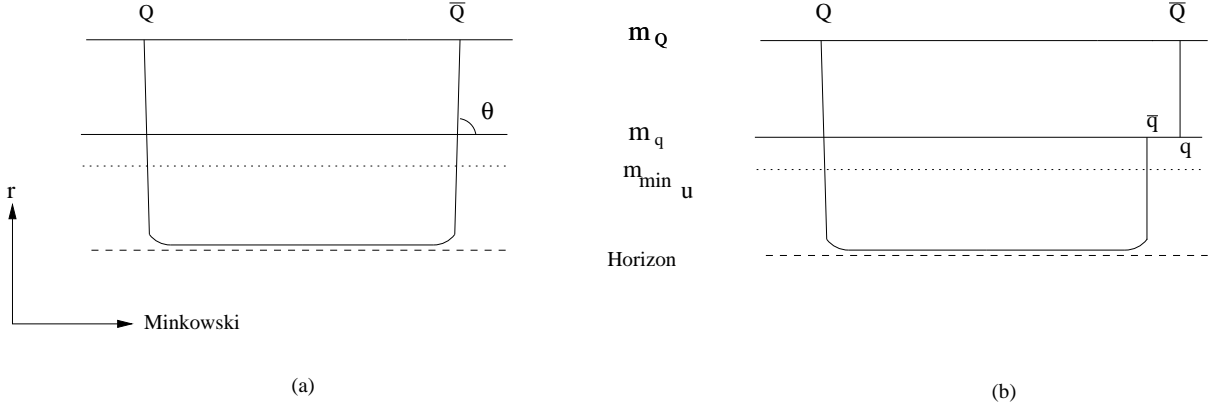}}
\caption{\small{(a) A large spin meson in Model I, bound state of two quarks of large mass $m_Q$, described by a string with both end-points on the same brane. The string intersects a second brane, corresponding to lighter quarks of mass $m_q$. (b) The strings after the splitting, representing two meson bound states of a heavy quark and a light quark.}}
\end{center}
\end{figure}

In the second model (that we will call Model II) one adds to the background $N_f\ll N_c$ D8/anti-D8 pairs \cite{ss}.
These branes reach the bottom of the space (minimal radius position), so they introduce massless flavors to the theory and nicely realize chiral symmetry breaking with the pions as Goldstone bosons.
The string describing the high spin meson spins on the D8 world-volume at the horizon, as can be seen in Figure 2 (a).

The process we want to study in Model I involves the introduction of a second flavor probe brane, corresponding to a lighter mass $m_q<m_Q$.
For the metric in the form:
\begin{equation}
ds^2=e^{A(r)}(-dt^2+d\rho^2+\rho^2d\eta^2+dx_3^2)+e^{B(r)}dr^2+G_{ij}d\phi^id\phi^j
\end{equation}
and spinning string configuration:
$t=\tau,\ \eta=\omega\tau,\ r=\sigma,\ \rho=\rho(\sigma),\ \phi^i=\phi^i_Q$,
the angle at which the string intersects the second brane is: 
\begin{equation}\label{angle}
\cos^2\theta=\frac{(\rho^\prime(r_q))^2}{e^{B(r_q)-A(r_q)}+(\rho^\prime(r_q))^2}.
\end{equation}

The crucial point now is that when $\theta\neq \frac{\pi}{2}$ there is a net transversal force, due to the string tension, in the direction of the brane world-volume, so that the string can split, describing the meson decay $\bar Q Q\rightarrow \bar Q q + \bar q Q$, see Figure 1 (b).
Note that the decay is highly constrained, since there are only two intersection points at which the splitting can happen, so there is  no phase space.

The process in Model II has instead a large phase space. 
The string can in fact split at any point, describing the decay $\bar q q\rightarrow \bar q q + \bar q q$, as in Figure 2 (b).
\begin{figure}
\begin{center}
\scalebox{0.7}{\includegraphics{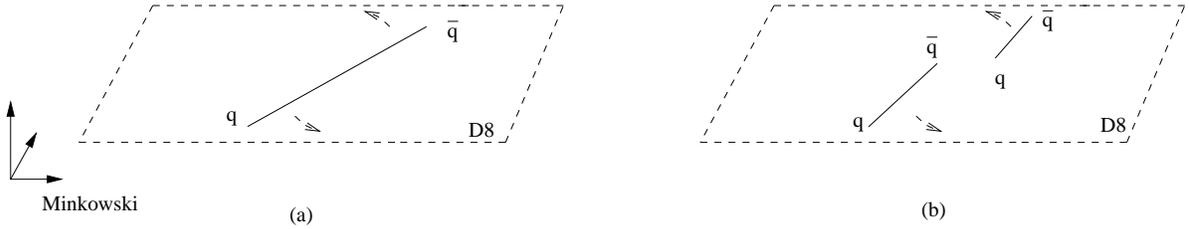}}
\caption{\small{(a) A large spin meson in Model II. (b) The strings after the splitting, representing two meson bound states of massless quarks.}}
\end{center}
\end{figure}
 
The question we are going to answer in the following is: what are the decay rates for these processes?
In order to perform such a computation we use some simplifications borrowed from \cite{jjp}. 
The most relevant one is that since the splitting process is a local one and the space is weakly curved, we can compute the rate in flat space, and eventually use the effective $\alpha^\prime_{eff}$ and dilaton that depend on the warp factor of the metric.
This consideration allows to reduce the computation to the rate for the splitting of a macroscopic string intersecting at generic angle (for Model I) or lying on (for Model II) a generic Dp-brane.
We do not report here the details of these computation, that can be found in \cite{noi1,noi2}, and just state and comment the results in the next section.

\section{The decay rates}

\subsection{The rates in flat space}

The decay rates in flat space read
\begin{equation}\label{gamma1}
\Gamma_{I} = \frac{g_s}{16\pi\sqrt{\alpha^\prime}}\cdot \frac{(2\pi\sqrt{\alpha^\prime})^{(8-p)}}{V_\perp} \cdot\frac{\cos^2\theta}{\sin\theta},
\end{equation}
\begin{equation}\label{gamma2}
\Gamma_{II} = \frac{g_s}{32\pi^2\alpha^\prime}\cdot\frac{(2\pi\sqrt{\alpha^\prime})^{(9-p)}}{V_\perp}\cdot L.
\end{equation}
A few comments are in order.
Apart from the obvious $g_s$ and dimensional ($\alpha'$) factors, in both formulas there is a suppression factor that increases with the volume of the directions transverse to the brane.
This effect is due to the quantum delocalization of the string and just means that the distance between the string and the brane must be of order $\sqrt{\alpha'}$ for the splitting to happen.
In the rate $\Gamma_{I}$ for the splitting of a string intersecting a Dp-brane (Model I) there is the expected angular dependence.
In fact, the $\sin\theta$ term tells that the probability of 
breaking increases as the string is more and more parallel to the brane, since the tension creates a bigger transversal force.
Instead, the $\cos^2\theta$ factor states that the rate vanishes for $\theta\rightarrow \pi/2$, since there is no transversal force. 
In the rate $\Gamma_{II}$ for the splitting of a string lying on a Dp-brane (Model II)
there is only the phase factor term: the rate is proportional to the length $L$ of the string, since it can split at any point.

\subsection{Preliminaries}

Before translating the formulas above in field theory language, let us introduce a bit of notation.
The background generated by the wrapped D4-branes reads
\begin{eqnarray}
ds^2=(\frac uR)^{3/2} (dx_\mu dx^\mu + \frac{4R^3}{9u_h}f(u)d\theta_2^2) + (\frac{R}{u})^{3/2}  \frac{du^2}{f(u)} + R^{3/2}u^{1/2} d\Omega_4^2,\nonumber\\ 
 f(u)=(u^3-u_h^3)/u^3, \qquad  \qquad \qquad\qquad e^{\Phi}=g_s\Bigl( \frac{u}{R}\Bigr)^{3/4},
\end{eqnarray}
where $u_h$ is the minimal radial position, the bottom of the space (the ``horizon'').
The String/Field theory dictionary allows to connect the string and field theory quantities \cite{bcmp}: 
$u_h=\frac{\lambda m_0 \alpha'}{3},\ g_s=\frac{\lambda}{3\pi N_c m_0 \sqrt{\alpha'}},\ R^3=\frac{\lambda\alpha'}{3m_0},\ T=\frac{\lambda m_0^2}{6\pi}$,
where $\lambda=g^2_{YM}N_c$ is the 't Hooft coupling at the UV cut-off, $m_0$ is the glueball and KK scale, and $T$ is the field theory string tension. 
Note that in this theory there are two distinct energy scales.

``High spin meson'' in both models refers to the regime $J\gg \lambda$. 

As a final preliminary, the quark mass in Model I is connected to the minimal radial position $u_Q$ of the brane by the relation
\begin{equation}\label{mass}
m_Q = {T\over m_0}\int_{1}^{u_Q/u_h} dz \left[1-{1\over 
z^3}\right]^{-{1\over2}}.
\end{equation} 

\subsection{Meson decay: Model II}

We now have all the ingredients to translate the formulas for the rates in flat space in the field theory context.
Let us begin by the simplest case, namely Model II, formula (\ref{gamma2}). 
As for the first factor, using the dictionary above we can immediately substitute:
$
\frac{g_s}{\alpha'} \rightarrow \frac{e^{\Phi}}{\alpha'_{eff}}=\frac{\lambda}{N_c}\frac{m_0^2\lambda^{3/2}}{3^{5/2}\pi}.
$
Moreover, since the spinning string is on the leading Regge trajectory, one has the very well known relations between the length, spin and energy: 
$
L=\sqrt{\frac{8 J}{\pi T}}=\frac{2M}{\pi T}, 
$
where $M$ is the meson mass.

Finally, as we already stressed, the string fluctuations create a broadening in the direction transverse to the D8. 
We can calculate this delocalization from the quadratic fluctuations of the corresponding massive world-sheet field \cite{jjp}.
The result is a logarithmic rather than power-like dependence on the size of the transverse dimension:
$
\frac{(2\pi \sqrt{\alpha'})^{(9-p)}}{V_{\perp}}= \frac{2\pi}{\log^{1/2}(1+\frac{8\pi T}{9m_0^2})}.
$

We can now write down the formula for the decay rate of a large spin meson made by massless quarks
\begin{equation}
\Gamma_{II}= \frac{\lambda}{N_c} \frac{1}{6\sqrt{2}\pi^{3/2}}\frac{1}{\log^{1/2}(1+\frac{8\pi T}{9m_0^2})}\frac{\sqrt{T}}{m_0}M.
\end{equation}
It obviously describes a $1/N_c$ process, increasing with the coupling $\lambda$.
Its most important feature is that the rate is linear in the mass $M$ of the meson.
Finally, the rate depends on the ratio of the two distinct mass scales of the theory, $\sqrt{T}$ and $m_0$; in a theory with only one mass scale $\Lambda_{QCD}$, such as QCD, one would expect that the rate would just read $\Gamma_{II}\sim \lambda M / N_c$.
Taking into account the rotation of the meson and the relativistic $\sqrt{1-v^2}$ factor, the rate in the laboratory reference frame is just the one above multiplied by a constant $\pi/4$ factor \cite{jjp}.

\subsection{Meson decay: Model I}

Let us now translate $\Gamma_{I}$ in (\ref{gamma1}).
As before, we have the coupling and dimensional factors:
$
\frac{g_s}{\sqrt{\alpha'}} \rightarrow \frac{e^{\Phi}}{\sqrt{\alpha'_{eff}}}=\frac{g_s}{\alpha'}(\frac{u_q}{R})^{3/4},
$
and the transverse direction volume:
$ 
V_{\perp}=2\pi R_{\theta_2} \cdot 2\pi R_{S^4}= \frac{8\pi^2u_q}{3u_h^{1/2}}R^{3/2}f^{1/2}(u_q).
$

We then need the slope of the string profile $\rho'(u)$ in order to calculate the intersection angle $\theta$ from (\ref{angle}). 
It turns out that there is an analytic expression of $\rho'(u)$ only in the large mass limit $m_Q \gg T/m_0$, when
the string profile is approximated by a corrected Wilson line spinning slowly \cite{pt}
\begin{equation}
\rho'(u)\approx {(Ru_h)^{3/2}\over u_h^3(x^3-1)}\left[1-{x^3(x-1)\over 
y(x^3-1)}\right], \qquad\qquad\qquad x\equiv u_q/u_h, \quad y\equiv 
u_Q/u_h.
\end{equation}
In the formula above we have used the fact that for $J \gg \lambda$ the lower point of the string profile is equal to $u_h$ up to exponentially (with $J$) suppressed terms:  $u_0\sim u_h(1+e^{-\frac{3m_0L}{2}})$, where $J$ is proportional to some power $\alpha$ of $L$, $J\sim L^{\alpha}$ \cite{pando}. 

Thus, the decay rate for a large spin meson made up of large mass quarks reads
\begin{equation}\label{dec}
\Gamma_I={\lambda m_0\over 16\pi^2N_c}{\sqrt{x}\over(x^3-1)}\left[1+ {1\over y}{(x-1)(1-2x^3)\over(x^3-1)}\right].
\end{equation}
Here we did not explicitate the variables $x,\ y$ because the formula (\ref{mass}) connecting these radial positions of the branes to the masses of the quarks gives some un-explicit special functions.
But we can have clear analytic formulas in the limit of large and small $m_q$.
In the large mass limit $m_q \gg T/m_0$
\begin{equation}\label{dec1}
\Gamma_I \sim \frac{\lambda}{16\pi^2 N_c} \left(\frac{T}{m_0}\right)^{5/2} \frac{m_0}{m_q^{5/2}}\left[1-2{m_q\over m_Q}\right],
\end{equation}
that in a ``one-scale limit'' would read $\Gamma_I \sim\frac{\lambda}{N_c}\frac{\Lambda_{QCD}^{7/2}}{m_q^{5/2}}\left[1-2{m_q\over m_Q}\right].$

In the small mass limit $x \approx x_{min} (\approx 1.04$ \cite{myers})
\begin{equation}\label{dec2}
\Gamma_I \sim {\lambda\over 36\pi^2N_c}\left({T\over m_0}\right)^2 
{m_0\over m_q^2}\left[1-{T\over 3m_0 m_Q}\right],
\end{equation}
that in a ``one-scale limit'' would read $\Gamma_I \sim {\lambda\over N_c} {\Lambda^3_{QCD}\over m_q^2}\left[1-{\Lambda_{QCD}\over m_Q}\right]$.

In the laboratory reference frame these rates must be multiplied by the relativistic $\sqrt{1-v^2}$ factor.
Since the decay can happen only around the heavy quarks one can approximate with $L/2$ the distance of the splitting point from the center of rotation.
Using the formula for the angular velocity from \cite{pt}, the relativistic factor reads: $\sqrt{\frac{2m_Q/L}{T+2m_Q/L}}$.
Remember that $L$ is proportional to some power of $J$ \cite{pando}.
 
Let us comment a bit on the formulas for the decay.
They describe of course a $1/N_c$ process that increases with the coupling $\lambda$.
Since the decay happens by pair-production of quarks of mass $m_q$, as the latter becomes smaller and smaller the decay is more and more probable.
Nevertheless, this suppression with $m_q$ is only power-like, so this is the leading decay channel in this ``dual of QCD'' at strong coupling: other processes involving instantonic world-sheet transitions are exponentially suppressed with $m_q$ and therefore are subleading. 
There is a very peculiar dependence on the decaying meson mass.
It is mild and only enters the formulas through the heavy quark mass.
The rate increases with $m_Q$ and goes to a constant for $m_Q \rightarrow \infty $.
Another relevant feature is that the decay is ``asymmetric'': one of the decay products has much larger spin and energy than the other, as can be seen from Figure 1 (b).
Moreover, the rate is dependent on the spin $J$ only through the relativistic factor and goes to zero in the laboratory reference frame as $J\rightarrow \infty$.
Finally, practically the same result applies to mesons made up of different heavy quarks. 

\section{Final considerations}

The physical picture that emerges for the decays is the following.
In Model I (decay of large spin, massive-quark mesons) the flux tube connecting the two quarks in the meson has almost constant energy density apart from a small region around the quarks (as can be seen from the shape of the string in figure 1 (a)).
In the decay to massive quarks, the flux tube has enough energy density for pair 
production of the light quarks only around the heavy quarks.
Thus, the tube can split only at these points.

Instead, in Model II (decay of large spin, massless-quark mesons) the flux tube has constant energy density everywhere. 
The decay is by pair-production of massless quarks, so every piece of the 
tube has enough energy for the process and as a result the rate is proportional to the mass (length) of the meson.

Unfortunately, it is quite difficult to check whether these results in our gauge theory models with string duals, in the planar limit and at strong coupling, could have some phenomenological relevance, because of lack of experimental data for large spin mesons.

Let us conclude with the observation that one can evaluate along the same lines described in this paper the meson decay rate in ${\cal N}=4$ SYM with flavors, and the decay rate of high spin glueballs dual to folded strings \cite{noi1}.
In the latter case the rate is particularly simple, $\Gamma \sim \frac{\lambda}{N^2}\frac{T^{5/2}}{m_0^4}$ ($\Gamma \sim \frac{\lambda}{N^2}\Lambda_{QCD}$ in the ``one-scale limit'').
It is a $1/N_c^2\ $ process that increases with the coupling, exhibiting no spin dependence.

\begin{center}    
{\large  {\bf Acknowledgments}}
\end{center}
This work is partially  supported by the European
Commission contracts MRTN-CT-2004-005104, MRTN-CT-2004-503369,
CYT FPA 2004-04582-C02-01, CIRIT GC 2001SGR-00065, MEIF-CT-2006-024173.

\end{document}